\documentclass{optica-article}

\journal{opticajournal} 

\articletype{Research Article}

\usepackage{lineno}

\newcommand{\rr}{\mathbf{r}}
\newcommand{\rp}{\mathbf{r}_{\perp}}
\newcommand{\jj}{\mathbf{j}}
\newcommand{\JJ}{\mathbf{J}}

\begin{document}

\title{Hidden transverse-flow topology in partially coherent structured light}

\author{Rosario Mart{\'\i}nez-Herrero and \'Angel S. Sanz\authormark{*}}

\address{Department of Optics, Faculty of Physical Sciences, Universidad Complutense de Madrid, Pza.\ Ciencias 1, Ciudad Universitaria -- 28040 Madrid, Spain}

\email{\authormark{*}a.s.sanz@fis.ucm.es} 


\begin{abstract*} 
Partial coherence is commonly viewed as a mechanism that reduces contrast or smooths intensity structure. Here we show that it can also encode hidden transverse-flow topology. Starting from the cross-spectral density, we formulate a generalized transverse flux, an effective velocity field, and the associated flux trajectories for quasi-monochromatic partially coherent paraxial beams. This construction converts two-point correlation phases into local transport information and reduces to the usual coherent energy-flow picture in the single-mode limit.
Because the generalized flux is obtained through a local differential operation on the cross-spectral density, the proposed trajectories are, in principle, reconstructible from sufficiently resolved phase-sensitive measurements of the complex second-order coherence function, subject to the regularization required by numerical differentiation.
Two analytical beam families expose this hidden topology. In twisted Gaussian Schell-model beams, a Gaussian intensity hides a distributed rotational flow with nonzero vorticity. In Laguerre-Christoffel-Darboux beams, sources with identical intensity profiles can have different flux topology, producing either spiral or purely radial trajectories. Thus, intensity and coherence magnitude do not exhaust the physical information contained in the cross-spectral density: partial coherence can reorganize the hidden topology of transverse optical transport.
\end{abstract*}



\section{Introduction}
\label{sec:intro}

Structured light has become a central topic in modern optics because the spatial amplitude, phase, polarization, and coherence properties of optical fields can be engineered~\cite{Forbes2021}. Such control has enabled beams with nontrivial propagation features, including self-acceleration~\cite{Berry1979,Siviloglou2007OL,Siviloglou2007PRL}, self-healing, orbital angular momentum (OAM), abrupt focusing, and tailored diffraction. In most practical situations, however, structured beams are not fully coherent. Their properties are affected by source fluctuations, finite apertures, partial spatial coherence, environmental perturbations, or deliberately engineered incoherence. In this regime, beams with similar, or even identical, intensity distributions may nevertheless transport optical energy in markedly different ways.

The standard theoretical descriptor of quasi-monochromatic partially coherent paraxial light is the cross-spectral density (CSD), which encodes the spatial correlations between pairs of points on a given transverse plane~\cite{MandelWolf1995,Wolf1982}. From the CSD one obtains the intensity distribution, the complex degree of coherence, and related coherence measures. These quantities provide essential information about where optical energy is distributed and how field values are correlated. However, they do not by themselves make explicit how optical energy is locally redistributed during propagation. In particular, the phase structure of the two-point correlation function may contain transverse-momentum information that is invisible in the intensity.

For fully coherent paraxial beams, this question can be addressed through the transverse energy flux~\cite{BerryMcDonald2008,Sanz-JOSAA:12,Sanz_ApplSci:20}. The ratio between the flux and the intensity defines an effective transverse velocity field, whose integral curves provide a trajectory representation of beam propagation. Such trajectories do not represent material particles; they are streamlines of the optical energy flow. In the coherent case, this construction is directly tied to the gradient of a single optical phase. It therefore provides a local, phase-sensitive view of propagation, complementary to the usual density plots of the intensity.

In this work we show that this transport information can be extracted directly from the CSD and used to reveal flow topology that is hidden at the intensity level. We formulate the corresponding generalized transverse flux, which satisfies a continuity equation for the averaged intensity~\cite{Sanz2025OLT}. This flux leads naturally to an effective transverse velocity field and to generalized flux trajectories.
The construction reduces to the usual coherent flux-trajectory picture when the CSD contains a single coherent mode, but it remains valid for physically admissible quasimonochromatic partially coherent beams whose CSD is sufficiently differentiable in a neighborhood of the coincidence set ${\bf r}_1={\bf r}_2$, whether they are represented by coherent modes or by ensembles of random field realizations.
In this sense, the CSD is not only a descriptor of intensity and coherence; it also defines an effective transverse transport field.

The conceptual point is that partial coherence should not be regarded only as a mechanism that reduces fringe visibility or smooths fine spatial details. It can also reorganize the topology of transverse energy flow.
Throughout this work, ``hidden'' means not inferable from the spectral density alone, rather than absent from the full CSD. We distinguish two related situations: intensity-hidden transport, in which the spectral density does not reveal a flow component encoded in the CSD phase, and intensity-degenerate transport inequivalence, in which distinct CSDs produce the same spectral density but different generalized fluxes. Generalized flux trajectories make both situations visible by converting two-point phase correlations into local transport information.
They reveal, for example, whether a given beam structure is fed or depleted by surrounding regions, whether rotational correlations persist in the absence of obvious intensity signatures, or whether two sources are intensity-equivalent but transport-inequivalent.

Recent developments have extended optical topology to spatiotemporal structured fields and have also revealed nontrivial structures encoded in the correlations of partially coherent light ~\cite{Liu2025STOVBursts,Liu2025ToroidalDynamics, Fan2026SpatiotemporalModes,Zhou2026LocalizedKnots, Liu2026SpatiotemporalCoherence,Zhou2026CoherenceToroidal}. The present work addresses a complementary problem: the local transverse transport topology extracted from the antisymmetric coincidence derivative of the CSD of a quasimonochromatic partially coherent field.

Within the more specific context of trajectory-based descriptions, previous analyses of Airy-type beams showed the usefulness of this approach for coherent and partially coherent structured fields~\cite{MartinezHerrero2022,Sanz2024OE,Sanz2025OLT}. In those cases, the emphasis was mainly on the persistence, degradation, or redistribution of recognizable propagation features such as self-acceleration and finite-energy truncation effects. Here the focus is different. We consider beam families in which the intensity can be smooth, rotationally symmetric, or even shared by distinct sources, while the underlying CSD phase structure produces different transverse-flow topologies. The aim is therefore not only to track visible beam features, but to uncover transport structures that are hidden at the intensity level.

\begin{figure*}[!t]
\centering
\includegraphics[width=\textwidth]{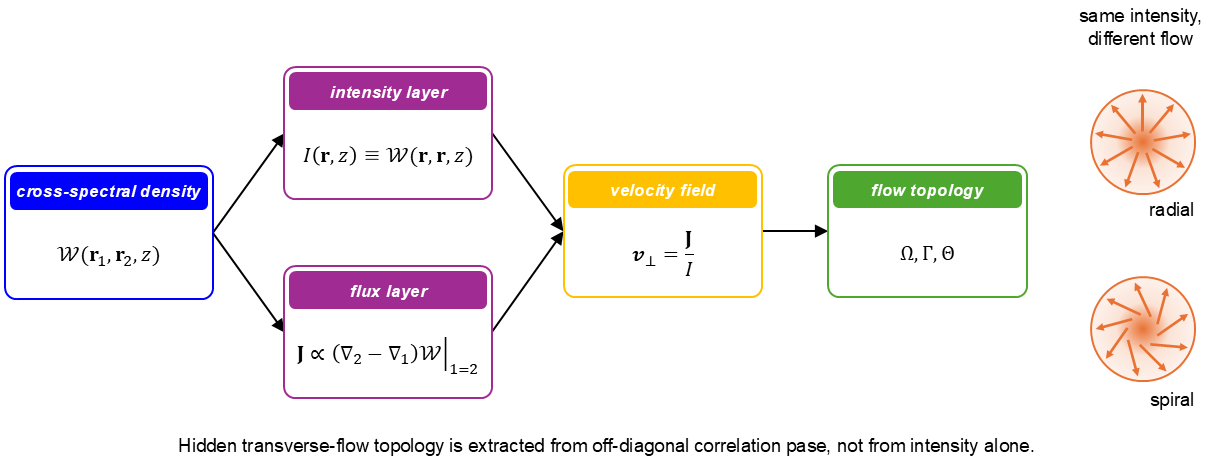}
\caption{
Conceptual structure of the generalized flux-trajectory formulation.
The cross-spectral density \(\mathcal{W}(\mathbf r_1,\mathbf r_2,z)\) contains both the diagonal intensity content and the off-diagonal correlation-phase information.
The former gives the spectral density \(I(\mathbf r,z) \equiv \mathcal{W}(\mathbf r,\mathbf r,z)\), whereas the latter gives access to the generalized transverse flux \(\mathbf J\).
Together, \(I\) and \(\mathbf J\) define the effective velocity field \(\mathbf v_\perp=\mathbf J/I\).
The topology of this velocity field is characterized through streamlines and diagnostics such as circulation \(\Gamma\), vorticity \(\Omega\), and accumulated angular displacement \(\Theta\).
The same intensity distribution can therefore support different transverse-flow topologies.
}
\label{fig:conceptual_pipeline}
\end{figure*}

The logic of the construction is summarized schematically in Fig.~\ref{fig:conceptual_pipeline}.
The CSD provides the intensity through its diagonal part, but it also contains off-diagonal phase-correlation information from which a generalized transverse flux can be extracted and, consequently, a hidden flow topology can be reconstructed.

The paper is organized as follows.
Section~\ref{sec:formalism} presents the generalized flux formulation for coherent and partially coherent paraxial beams. Section~\ref{sec:diagnostics} in\-tro\-du\-ces trajectory-based diagnostics aimed at characterizing transverse-flow topology. Section~\ref{sec:examples} develops two analytical examples: twisted Gaussian Schell-model (TGSM) beams and Laguerre-Christoffel-Darboux (LCD) beams. These examples show, respectively, how a nonseparable coherence phase generates spiral transport in a beam with Gaussian intensity, and how sources with the same intensity may display different flow topology because of their angular coherence structure. Section~\ref{sec:discussion} discusses the physical interpretation and possible implications. Section~\ref{sec:conclusions} summarizes the conclusions.


\section{Generalized flux formulation}
\label{sec:formalism}

The aim of this section is to identify the local transport structure encoded in a partially coherent paraxial field. We first recall the usual flux formulation for coherent beams, where the transverse flow is governed by the gradient of a single optical phase. We then show that, even when such a phase is no longer defined, the CSD contains enough two-point phase information to define a generalized transverse flux. This flux is the central quantity from which the hidden transverse-flow topology will be extracted.


\subsection{Coherent paraxial beams}
\label{sec:2.1}

Let us consider a monochromatic scalar paraxial beam propagating along the longitudinal coordinate $z$. The optical field is written as
\begin{equation}
\label{eq:coherent_field}
\Psi(\rp,z)=\psi(\rp,z)\exp(ikz).
\end{equation}
Here ${\bf r}_\perp=(x,y)$ denotes the transverse position and $k$ is the wavenumber in the medium.
We adopt the time-harmonic convention $\exp(-i\omega t)$, so that Eq.~\eqref{eq:coherent_field} represents a forward-propagating carrier $\exp(ikz)$. With this convention, the slowly varying amplitude satisfies the paraxial Helmholtz equation
\begin{equation}
\label{eq:paraxial}
i\frac{\partial \psi}{\partial z}=-\frac{1}{2k}\nabla_{\perp}^{2}\psi.
\end{equation}
Equation~\eqref{eq:paraxial} implies the transverse continuity equation
\begin{equation}
\label{eq:coherent_continuity}
\frac{\partial I}{\partial z}=-\nabla_{\perp}\cdot \jj ,
\end{equation}
where
\begin{equation}
\label{eq:coherent_intensity}
I(\rp,z)=|\psi(\rp,z)|^{2}
\end{equation}
is the coherent intensity, and
\begin{equation}
\label{eq:coherent_flux}
\jj(\rp,z)=\frac{1}{2ik}\left(\psi^{*}\nabla_{\perp}\psi-\psi\nabla_{\perp}\psi^{*}\right)
\end{equation}
is the transverse flux.
The corresponding effective transverse velocity field is obtained as the ratio between the transverse flux and the intensity~\cite{Sanz2025OLT},
\begin{equation}
\label{eq:coherent_velocity}
\mathbf{v}_{\perp}(\rp,z)=\frac{\jj(\rp,z)}{I(\rp,z)}.
\end{equation}
Here, ``velocity'' denotes the transverse displacement rate $d{\bf r}/dz$ with respect to the propagation coordinate, rather than a material velocity defined with respect to time.
If the field amplitude is written in polar form,
\begin{equation}
\label{eq:polar_form}
\psi(\rp,z)=A(\rp,z)\exp[iS(\rp,z)],
\end{equation}
then Eq.~\eqref{eq:coherent_velocity} becomes
\begin{equation}
\label{eq:phase_velocity}
\mathbf{v}_{\perp}(\rp,z)=\frac{1}{k}\nabla_{\perp}S(\rp,z).
\end{equation}
Thus, in the coherent case, the transverse-flow topology is determined by the spatial organization of a single phase function. Vortices, caustic-like structures, separatrices, or spiral streamlines can therefore be traced back to the phase-gradient field associated with the coherent amplitude.

The associated coherent flux trajectories are obtained from
\begin{equation}
\label{eq:coherent_trajectories}
\frac{d\rp}{dz}=\mathbf{v}_{\perp}(\rp,z).
\end{equation}
When the initial conditions are sampled according to the input intensity, the density of trajectories at later planes is consistent with the propagated intensity distribution.

\subsection{Partially coherent beams and the CSD}

For compactness, in what follows \(\mathbf r\) denotes a transverse coordinate.
For a quasi-monochromatic partially coherent beam, the cross-spectral density is defined as
\begin{equation}
\label{eq:csd_definition}
\mathcal{W}(\rr_{1},\rr_{2},z)=\langle U^{*}(\rr_{1},z)U(\rr_{2},z)\rangle.
\end{equation}
The averaged intensity is obtained from the diagonal elements,
\begin{equation}
\label{eq:csd_intensity}
I(\rr,z)=\mathcal{W}(\rr,\rr,z).
\end{equation}
Note that, strictly speaking, this quantity represents the spectral density of the quasimonochromatic field. In the following, we occasionally use the term ``intensity'' as a convenient shorthand when no ambiguity can arise.

A physically admissible CSD must define a positive-semidefinite correlation kernel.
It can therefore be written in coherent-mode form as~\cite{MandelWolf1995,MartinezHerrero1979}
\begin{equation}
\label{eq:coherent_mode_expansion}
\mathcal{W}(\rr_{1},\rr_{2},z)=\sum_{n}p_{n}\phi_{n}^{*}(\rr_{1},z)\phi_{n}(\rr_{2},z),\qquad p_{n}\ge 0.
\end{equation}
Each mode satisfies the paraxial equation and carries its own transverse flux \cite{Sanz2025OLT},
\begin{equation}
\label{eq:modal_flux}
\jj_{n}=\frac{1}{2ik}\left(\phi_{n}^{*}\nabla_{\perp}\phi_{n}-\phi_{n}\nabla_{\perp}\phi_{n}^{*}\right).
\end{equation}
The averaged intensity then satisfies the continuity equation
\begin{equation}
\label{eq:continuity_equation}
\frac{\partial I}{\partial z}
=
-\nabla_\perp\cdot\mathbf J ,
\end{equation}
with the generalized transverse flux
\begin{equation}
\label{eq:generalized_flux_modes}
\JJ(\rr,z)=\sum_{n}p_{n}\jj_{n}(\rr,z).
\end{equation}
In the following, we restrict the pointwise construction to physically admissible CSDs that are continuously differentiable with respect to both transverse arguments in a neighborhood of the coincidence set ${\bf r}_1={\bf r}_2$. Under this sufficient regularity condition, the antisymmetric coincidence derivative entering the generalized transverse flux is well defined.

Thus, equivalently, the same flux can be written directly in terms of the CSD as
\begin{equation}
\label{eq:generalized_flux_csd}
\JJ(\rr,z)=\frac{1}{2ik}\left(\nabla_{2}-\nabla_{1}\right)\mathcal{W}(\rr_{1},\rr_{2},z)\bigg|_{\rr_{1}=\rr_{2}=\rr},
\end{equation}
where $\nabla_1$ and $\nabla_2$ denote transverse gradients with respect to the first and second spatial arguments, ${\bf r}_1$ and ${\bf r}_2$, respectively.
Equation~\eqref{eq:generalized_flux_csd} is the central object of the present formulation. It extracts the antisymmetric phase-gradient content of the two-point correlation function at coincidence points. In doing so, it converts correlation-phase information into a local transverse flux. The CSD therefore encodes not only the intensity and the degree of coherence, but also the transport field that determines how the averaged optical energy is redistributed across the transverse plane.
In the coherent-mode representation, the same pointwise construction also assumes that the weighted sum of modal currents and the corresponding derivative series converge locally, so that differentiation and modal summation may be interchanged.


\subsection{Generalized effective velocity and trajectories}

The generalized transverse effective velocity field is defined as \cite{Sanz2025OLT}
\begin{equation}
\label{eq:generalized_velocity}
\mathbf v_{\perp}(\mathbf r,z)=\frac{\mathbf J(\mathbf r,z)}{I(\mathbf r,z)}.
\end{equation}
The effective velocity is defined pointwise on the subset of the observation plane where $I({\bf r},z)>0$. At zeros of the spectral density, the flux may remain well defined, but the ratio ${\bf J}_\perp/I$ is generally undefined or singular. Such points must therefore be excluded from the regular trajectory domain or treated through an appropriate limiting or regularization procedure.

Within any region where ${\bf v}_\perp({\bf r},z)$ is regular, the generalized flux trajectories are obtained from
\begin{equation}
\label{eq:generalized_trajectories}
\frac{d\mathbf r}{dz}=\mathbf v_{\perp}(\mathbf r,z).
\end{equation}
These curves are $z$-parametrized trajectories of the averaged transverse transport field. They describe how the averaged optical energy of the partially coherent beam is locally redistributed during propagation and reduce to the usual coherent flux trajectories when the CSD contains a single coherent mode.
These propagation trajectories should be distinguished from the fixed-plane streamlines of the velocity field, obtained by fixing $z=z_\ast$ and integrating the frozen field ${\bf v}_\perp({\bf r},z_\ast)$ with respect to an auxiliary curve parameter. The displayed extent of such fixed-plane streamlines is determined by the selected seed points, plotting domain, and numerical termination criteria and therefore has no independent physical meaning (see numerical simulations shown in Sec.~\ref{sec:examples}).

An equivalent expression can be obtained if the partially coherent beam is represented as an ensemble of random realizations. In that case, the CSD is
\begin{equation}
\label{eq:ensemble_csd}
\mathcal{W}(\mathbf r_1,\mathbf r_2,z)=\langle \psi^*(\mathbf r_1,z)\psi(\mathbf r_2,z)\rangle,
\end{equation}
and the averaged intensity is
\begin{equation}
\label{eq:ensemble_intensity}
I(\mathbf r,z)=\langle |\psi(\mathbf r,z)|^2\rangle.
\end{equation}
The generalized flux is then
\begin{equation}
\label{eq:ensemble_flux}
\mathbf J(\mathbf r,z)=\frac{1}{2ik}
\left\langle
\psi^*\nabla_\perp\psi-\psi\nabla_\perp\psi^*
\right\rangle.
\end{equation}
Thus, the relevant velocity field is the ratio between the ensemble-averaged transverse flux and the ensemble-averaged intensity. It is not the ensemble average of the individual velocity fields. This distinction is essential whenever different coherent components or random realizations carry competing local momenta, because cancellations in the averaged flux may reorganize the topology of the resulting transport field.


\subsection{Physical interpretation}
\label{sec:IID}

The coherent velocity field is governed by the phase gradient of a single complex amplitude. In the partially coherent case, there is no unique global phase from which a velocity field could be obtained. Nevertheless, the CSD contains phase information of a different kind: a two-point correlation phase. Equation~\eqref{eq:generalized_flux_csd} shows that the local variation of this correlation phase at coincidence points defines a transverse flux.

The generalized trajectories therefore provide a way to visualize and quantify the effective transport dynamics encoded in the CSD. They are especially useful when different coherent modes or random field realizations contribute differently to the local transverse momentum.
In such cases, the intensity may be smooth, featureless, or shared by distinct sources, while the velocity field may contain radial expansion, circulation, shear, stagnation points, separatrices, or spiral regions.
The hidden transverse-flow topology is precisely the structure of the generalized velocity field that cannot be inferred from the spectral density alone.
The TGSM example below illustrates intensity-hidden transport: its Gaussian spectral density does not reveal the twist-induced azimuthal flow. The LCD comparison illustrates the stronger case of intensity-degenerate transport inequivalence: distinct CSDs have the same spectral density but different generalized fluxes.

The formulation is also formally analogous to the trajectory representation associated with reduced density matrices in quantum mechanics~\cite{Sanz_EPJD:07,Sanz_CPL:09}. In that analogy, the CSD plays the role of a density matrix in the position representation, and the generalized flux corresponds to the current associated with a mixed or reduced state. This reinforces the view of the CSD flux as a correlation-induced transport quantity rather than as a simple average of coherent phase gradients.

The generalized-flux construction is not restricted to a particular degree of coherence. It applies from the fully coherent single-mode limit to partially coherent multimode fields, provided that the CSD is physically admissible and sufficiently differentiable at coincidence.
The effect of coherence on the transverse flow cannot, however, be described by a single global coherence parameter.
The local transport field is determined by the antisymmetric coincidence derivative of the complex CSD, and therefore depends on both the modal weights and the correlation phases. Consequently, a change in coherence may preserve, weaken, enhance, reorganize, or cancel a given flow component, depending on how the underlying correlation structure is modified. The two beam families considered in Sec.~\ref{sec:examples} illustrate different possibilities: in TGSM beams, the physical admissibility condition couples the allowed twist to the transverse coherence length, whereas in the single-charge LCD family the incoherent addition of radial modes modifies the spectral-density profile without changing the normalized azimuthal velocity. By contrast, an equal-weight mixture of opposite topological charges cancels the azimuthal flux altogether.


\section{Diagnostics of hidden transverse-flow topology}
\label{sec:diagnostics}

The generalized velocity field defined by Eq.~\eqref{eq:generalized_velocity} contains more information than the intensity distribution alone. In particular, two partially coherent beams may have the same spectral density \(I(\mathbf r,z)\) but different generalized fluxes \(\mathbf J(\mathbf r,z)\), and therefore different transverse-flow topology.
In this section we introduce a minimal set of trajectory-based diagnostics that will be used to distinguish radial, spiral, and circulation-carrying transport structures.
They will be used below to interpret the TGSM and LCD examples.

\subsection{Radial and azimuthal transport}

For rotationally symmetric or nearly rotationally symmetric beams it is useful to decompose the generalized velocity field as
\begin{equation}
\label{eq:radial_azimuthal_decomposition}
\mathbf v_{\perp}(r,\varphi,z)
=
v_r(r,\varphi,z)\,\hat{\mathbf r}
+
v_\varphi(r,\varphi,z)\,\hat{\boldsymbol\varphi}.
\end{equation}
The radial component \(v_r\) accounts for diffractive expansion or contraction, whereas the azimuthal component \(v_\varphi\) accounts for circulation around the propagation axis. A beam whose intensity is rotationally symmetric may still have \(v_\varphi\neq 0\). In that case, the rotational structure is hidden at the intensity level but becomes apparent in the generalized flux.

For the analytical examples considered below, this decomposition provides a direct way to distinguish purely radial transport from transport with an azimuthal component, which gives rise to spiral propagation trajectories.
If \(v_\varphi=0\), the trajectories remain radial. If \(v_\varphi\neq 0\), the trajectories wind around the propagation axis, even if the intensity remains circularly symmetric.

\subsection{Divergence, vorticity, and circulation}

Local spreading or compression of neighboring trajectories is characterized by the divergence
\begin{equation}
\label{eq:flow_divergence}
D(\mathbf r,z) \equiv \nabla_{\perp}\cdot \mathbf v_{\perp}(\mathbf r,z).
\end{equation}
For a purely radial diffractive expansion, \(D\) measures the local dilution of the averaged optical energy flow. However, divergence alone does not characterize rotational transport. For two-dimensional transverse fields, the rotational content is measured by the scalar vorticity
\begin{equation}
\label{eq:flow_vorticity}
\Omega(\mathbf r,z)\equiv
\left[
\nabla_{\perp}\times \mathbf v_{\perp}(\mathbf r,z)
\right]_z,
\end{equation}
or, equivalently, by the circulation around a closed contour \(C\),
\begin{equation}
\label{eq:flow_circulation_contour}
\Gamma_C(z)\equiv
\oint_C \mathbf v_{\perp}\cdot d\boldsymbol{\ell}.
\end{equation}
A nonzero circulation signals the presence of azimuthal transport. This is particularly useful when the intensity does not display any visible rotational feature. In such cases, circulation and vorticity provide direct diagnostics of a hidden transverse-flow topology.

For an axially symmetric velocity field,
\begin{equation}
\label{eq:axial_velocity_decomposition}
\mathbf v_{\perp}(r,z)=v_r(r,z)\,\hat{\mathbf r}
+v_\varphi(r,z)\,\hat{\boldsymbol\varphi},
\end{equation}
the vorticity is
\begin{equation}
\label{eq:axial_vorticity}
\Omega(r,z)=\frac{1}{r}\frac{\partial}{\partial r}
\left[r v_\varphi(r,z)\right] ,
\end{equation}
and the circulation around a circle of radius \(r\) is
\begin{equation}
\label{eq:axial_circulation}
\Gamma(r,z)=2\pi r\,v_\varphi(r,z) .
\end{equation}
These two quantities distinguish, for example, the distributed vorticity of a TGSM beam from the radius-independent circulation associated with the angular phase structure of a Laguerre-type flow.

\subsection{Accumulated angular displacement}

A complementary global diagnostic is the angular displacement accumulated by a trajectory,
\begin{equation}
\label{eq:accumulated_angular_displacement}
\Theta(z;r_0)=
\varphi(z)-\varphi(0)
=
\int_0^z
\frac{v_\varphi[r(z'),z']}{r(z')}
\,dz' .
\end{equation}
This quantity measures the net rotation about the propagation axis experienced by a trajectory between the input plane and the plane \(z\).
In beams with hidden azimuthal transport, \(\Theta\) can be nonzero even when the intensity remains rotationally symmetric during propagation. It therefore provides a simple scalar measure of the strength and handedness of the hidden flow.

\subsection{Flow inequivalence for equal intensities}

The central situation considered in this work occurs when two partially coherent sources have the same intensity distribution but different generalized fluxes. Let two beams \(A\) and \(B\) satisfy
\begin{equation}
\label{eq:equal_intensities}
I_A(\mathbf r,z)=I_B(\mathbf r,z) =I(\mathbf r,z),
\end{equation}
but
\begin{equation}
\label{eq:unequal_fluxes}
\mathbf J_A(\mathbf r,z)\neq \mathbf J_B(\mathbf r,z).
\end{equation}
Then their generalized velocity fields are different,
\begin{equation}
\label{eq:unequal_velocities}
\mathbf v_A(\mathbf r,z)\neq \mathbf v_B(\mathbf r,z),
\end{equation}
and the beams are dynamically inequivalent despite being indistinguishable at the intensity level. A useful measure of this inequivalence is the intensity-weighted flow contrast
\begin{equation}
\label{eq:flow_contrast}
\mathcal C_{AB}(z) \equiv \sqrt{ \frac{\displaystyle \int I(\mathbf r,z)
\left| \mathbf v_A(\mathbf r,z)-\mathbf v_B(\mathbf r,z) \right|^2 d^2\mathbf r}{\displaystyle \int I(\mathbf r,z) \left[ |\mathbf v_A(\mathbf r,z)|^2 + |\mathbf v_B(\mathbf r,z)|^2 \right] d^2\mathbf r } } .
\end{equation}
This quantity is defined for fields sharing the same spectral density, so that the intensity distribution $I({\bf r},z)$ provides an unambiguous weighting measure. If the two fields have different intensities or different supports, a more general contrast would require the prior choice of a common nonnegative weight over the union of their supports and would therefore no longer be unique.

Note that Eq.~\eqref{eq:flow_contrast} vanishes only when the two velocity fields coincide almost everywhere within the support of the intensity. It is therefore sensitive to hidden differences in the CSD phase structure. In the analytical examples below, the inequivalence can already be identified through the azimuthal component of the velocity field and the circulation. The contrast \(\mathcal C_{AB}\) provides a corresponding global measure when a single scalar comparison between two transport fields is desired. In the LCD example below, the single-charge and balanced opposite-charge sources may share the same intensity profile, but their azimuthal fluxes, circulations, and trajectory topology are different.



\section{Hidden flow topology in analytical beam families}
\label{sec:examples}

We now apply the generalized flux formulation to two analytically tractable families of partially coherent structured beams. The purpose is not to introduce new sources, but to show explicitly how the same intensity-based description may hide different transverse-flow structures. The two examples are complementary. In TGSM beams, the intensity remains a circular Gaussian during propagation, while the nonseparable coherence phase generates a distributed azimuthal flow.
In LCD beams, two sources can share the same radial intensity profile but display different transverse-flow topology depending on their angular coherence structure.
These cases therefore provide direct demonstrations of the hidden transverse-flow topology encoded in the CSD.

\subsection{Twisted Gaussian Schell-model beam}

The TGSM beam provides a minimal example of hidden flow topology. Its spectral density remains Gaussian and circularly symmetric during propagation, but its CSD contains a nonseparable two-point phase that generates azimuthal transport. Thus, the twist is not primarily an intensity effect; it is a correlation-phase effect that becomes visible through the generalized flux.

TGSM beams were introduced by Simon and Mukunda as rotationally invariant Gaussian CSDs with a nonseparable twist phase~\cite{SimonMukunda1993}. Their physical interpretation and experimental demonstration were discussed by Friberg, Tervonen, and Turunen~\cite{Friberg1994}. Superposition viewpoints were developed by Ambrosini, Bagini, Gori, and Santarsiero~\cite{Ambrosini1994}, and a connection with partially coherent modified Bessel-Gauss beams was later given by Gori and Santarsiero~\cite{GoriSantarsiero2015}. Here we use the TGSM beam as an analytically solvable benchmark showing how rotational transport can be hidden in a beam whose intensity contains no visible rotational signature.
Recent experimental work has also shown that TGSM beams with controllable twist phase can be synthesized and characterized in terms of their OAM~\cite{Wang2022Nanophotonics}.

Let
\begin{equation}
\label{eq:tgsm_coordinates}
\mathbf r_j = \begin{pmatrix} x_j \\ y_j \end{pmatrix} ,
\qquad
\boldsymbol{\epsilon} =
\begin{pmatrix}
0 & 1\\
-1 & 0
\end{pmatrix}.
\end{equation}
Then
\begin{equation}
\label{eq:tgsm_cross_product}
\mathbf r_1^T\boldsymbol{\epsilon}\mathbf r_2 = x_1y_2-y_1x_2 .
\end{equation}
In the sign convention used here, the TGSM cross-spectral density at a plane \(z\) is written as
\begin{equation}
\label{eq:tgsm_csd}
\mathcal{W}(\mathbf r_1,\mathbf r_2,z) =
\frac{P}{2\pi w^2(z)}
\exp\left[\Xi_z(\mathbf r_1,\mathbf r_2)\right].
\end{equation}
The argument of the exponential function is defined as
\begin{equation}
\label{eq:tgsm_exponent}
\Xi_z ({\bf r}_1,{\bf r}_2) \equiv
-\frac{\mathbf r_1^2+\mathbf r_2^2}{4 w^2(z)}
-\frac{|\mathbf r_1-\mathbf r_2|^2}{2\delta^2(z)}
-\frac{ik(\mathbf r_1^2-\mathbf r_2^2)}{2R(z)}
+ik\mu(z)\mathbf r_1^T\boldsymbol{\epsilon}\mathbf r_2 .
\end{equation}
Here \(P\) is the total spectral power, \( w(z)\) is the beam-size parameter, \(\delta(z)\) is the transverse coherence length, \(R(z)\) is the phase-front radius of curvature, and \(\mu(z)\) is the twist parameter. The final term in Eq.~\eqref{eq:tgsm_exponent} is not a separable phase of the form \(\Phi(\mathbf r_2)-\Phi(\mathbf r_1)\). It is a genuine two-point phase associated with partial coherence.

Assume that \(z=0\) is the waist plane. The propagated TGSM beam keeps the form of Eq.~\eqref{eq:tgsm_csd}, with
\begin{equation}
\label{eq:tgsm_sigma_s}
 w(z)
=
 w_0
\left(
1+\frac{z^2}{\Lambda_0^2}
\right)^{1/2},
\end{equation}
\begin{equation}
\label{eq:tgsm_sigma_g}
\delta(z)
=
\delta_0
\left(
1+\frac{z^2}{\Lambda_0^2}
\right)^{1/2},
\end{equation}
\begin{equation}
\label{eq:tgsm_mu_z}
\mu(z)
=
\frac{\mu_0}{1+z^2/\Lambda_0^2},
\end{equation}
and
\begin{equation}
\label{eq:tgsm_radius_curvature}
R(z)
=
z+\frac{\Lambda_0^2}{z}.
\end{equation}
The TGSM Rayleigh-like parameter satisfies
\begin{equation}
\label{eq:tgsm_rayleigh_parameter}
\Lambda_0^{-2}
=
\frac{1}{4k^2 w_0^4}
\left(
1+\frac{4 w_0^2}{\delta_0^2}
\right)
+
\mu_0^2 .
\end{equation}
The twist increases the effective angular divergence through the additional term \(\mu_0^2\). Positivity of the CSD imposes the bound
\begin{equation}
\label{eq:tgsm_positivity_bound}
\delta^2(z)|\mu(z)|\le \frac{1}{k}.
\end{equation}
Equation~\eqref{eq:tgsm_positivity_bound} is the positive-semidefiniteness condition for the TGSM cross-spectral-density kernel and therefore ensures its physical admissibility. This condition explicitly couples the twist strength to the transverse coherence length, since $|\mu(z)|\leq 1/k\delta^2(z)$. In particular, for fixed nonzero $\mu(z)$, the limit $\delta(z)\to\infty$ is not admissible. The TGSM family thus provides a direct example of the general observation made in Sec.~\ref{sec:IID}: changing the coherence properties can constrain the magnitude of the physically accessible rotational flow.

The spectral density is obtained by setting \(\mathbf r_1=\mathbf r_2=\mathbf r\):
\begin{equation}
\label{eq:tgsm_intensity}
S(\mathbf r,z)
=
\mathcal{W}(\mathbf r,\mathbf r,z)
=
\frac{P}{2\pi w^2(z)}
\exp\left[
-\frac{\mathbf r^2}{2 w^2(z)}
\right].
\end{equation}
Equation~\eqref{eq:tgsm_intensity} shows that the spectral density is completely insensitive to the sign and magnitude of the twist parameter, except through the propagated beam width. At the intensity level, the beam is a circular Gaussian. The hidden structure appears only when the two-point phase of the CSD is converted into a local flux. Direct differentiation of Eq.~\eqref{eq:tgsm_csd} using Eq.~\eqref{eq:generalized_flux_csd} gives
\begin{equation}
\label{eq:tgsm_flux}
\frac{\mathbf J_\perp}{S}
=
\frac{\mathbf r}{R(z)}
+
\mu(z) \left( \hat{\mathbf z}\times\mathbf r \right) .
\end{equation}
The first term is the radial diffractive expansion associated with phase-front curvature. The second term is purely azimuthal and is produced solely by the nonseparable twist phase. Therefore,
\begin{equation}
\label{eq:tgsm_velocity_components}
v_r(r,z)=\frac{r}{R(z)},
\qquad
v_\varphi(r,z)=\mu(z)r .
\end{equation}
The azimuthal component has the form of a rigid-body-like rotation.
Because $v_\varphi=r\,d\varphi/dz$, the angular rate is $d\varphi/dz=\mu(z)$, which is therefore independent of the launch radius.
Thus, although the normalized velocity grows linearly with $r$, the corresponding flux, $J_\varphi=I\mu(z)r$, vanishes in the large-$r$ tails because of the Gaussian decay of the TGSM spectral density.
Accordingly, the flux and its intensity-weighted moments remain finite. The formal growth of $v_\varphi$ in the exponentially weak tails lies outside the physically relevant paraxial region and does not imply either unbounded energy transport or a causality violation.
The corresponding circulation around a circle of radius \(r\) is
\begin{equation}
\label{eq:tgsm_circulation}
\Gamma_{\rm TGSM}(r,z)
=
2\pi r\,v_\varphi(r,z)
=
2\pi \mu(z)r^2,
\end{equation}
and the vorticity is
\begin{equation}
\label{eq:tgsm_vorticity}
\Omega_{\rm TGSM}(r,z)
=
\frac{1}{r}
\frac{\partial}{\partial r}
\left[
r v_\varphi(r,z)
\right]
=
2\mu(z).
\end{equation}

\begin{figure*}[!t]
\centering
\includegraphics[width=\textwidth]{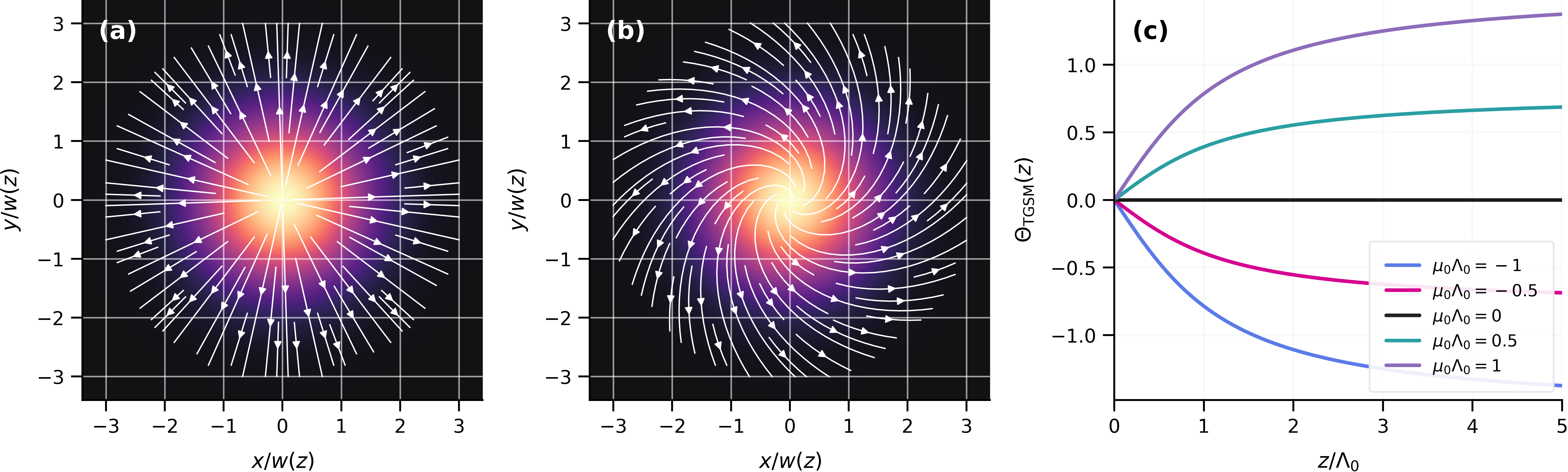}
\caption{
Hidden rotational flow in TGSM beams. The background color represents the normalized Gaussian spectral density $S(r,z)$ at the propagation plane $z/\Lambda_0=1$, shown in coordinates scaled by the propagated beam size $w(z)$.
White curves in panels (a) and (b) are fixed-plane streamlines of the generalized transverse velocity field at $z/\Lambda_0=1$.
They are integral curves of the frozen velocity field, rather than propagation trajectories integrated over a finite $z$ interval. Their displayed extent is set by the common seed distribution, plotting window, and numerical termination criteria and has no independent physical meaning.
(a) In the untwisted case, $\mu_0\Lambda_0=0$, and the flow is purely radial.
(b) For a twisted TGSM beam with $\mu_0\Lambda_0 \rightarrow 1^{-}$, one has $\mu(z)\Lambda_0=1/2$ at the displayed plane
(Note that the value $|\mu_0\Lambda_0|=1$ represents an asymptotic upper limit and is not attained for finite regular TGSM parameters). The spectral density remains Gaussian, but the azimuthal component $v_\varphi=\mu(z)r$ produces spiral fixed-plane streamlines.
The same seed distribution and plotting conventions are used in panels (a) and (b).
Streamlines at other propagation planes follow from the same expressions with the corresponding values of $R(z)$ and $\mu(z)$; their accumulated angular evolution is summarized in panel (c).
(c) Unlike the fixed-plane streamlines in panels (a) and (b), this panel shows the angular displacement accumulated by the propagated trajectories from the waist plane to $z$, $\Theta_{\rm TGSM}(z)=\mu_0\Lambda_0\arctan(z/\Lambda_0)$, for several values of $\mu_0\Lambda_0$.
The same twist parameter determines the vorticity and circulation, $\Omega_{\rm TGSM}=2\mu(z)$ and $\Gamma_{\rm TGSM}=2\pi\mu(z)r^2$.
}
\label{fig:tgsm_hidden_flow}
\end{figure*}

This hidden rotational structure is illustrated in Figs.~\ref{fig:tgsm_hidden_flow}(a) and \ref{fig:tgsm_hidden_flow}(b) through fixed-plane streamlines of the transverse velocity field at $z/\Lambda_0=1$.
These fixed-plane streamlines visualize the local organization of the flow and should not be confused with the $z$-parametrized propagation trajectories derived below.
At a different propagation plane, the corresponding fixed-plane streamline pattern is obtained from the same velocity field by evaluating $R(z)$ and $\mu(z)$ at the corresponding value of $z$. Thus, propagation changes the relative radial and azimuthal contributions continuously, while the accumulated rotation over the complete propagation interval is shown in Fig.~\ref{fig:tgsm_hidden_flow}(c).
In other words, a TGSM beam may have a featureless Gaussian intensity while carrying a distributed rotational flow with uniform vorticity. The handedness of this hidden topology is fixed by the sign of \(\mu(z)\).

The trajectory equations are thus obtained from the integration of the equations
\begin{equation}
\label{eq:tgsm_radial_equation}
\frac{dr}{dz}
=
\frac{r}{R(z)},
\end{equation}
and
\begin{equation}
\label{eq:tgsm_angular_equation}
\frac{d\varphi}{dz}
=
\mu(z).
\end{equation}
Substituting Eqs.~\eqref{eq:tgsm_mu_z} and \eqref{eq:tgsm_radius_curvature} in the above equations, they can readily be integrated in a closed form.
Specifically, for \(r(0)=r_0\) and \(\varphi(0)=\varphi_0\), one obtains
\begin{equation}
\label{eq:tgsm_radial_solution}
r(z) = r_0 \sqrt{ 1+\frac{z^2}{\Lambda_0^2} } ,
\end{equation}
and
\begin{equation}
\label{eq:tgsm_angular_solution}
\varphi(z) = \varphi_0 + \mu_0\Lambda_0
 \arctan  \left( \frac{z}{\Lambda_0} \right).
\end{equation}
The accumulated angular displacement,
\begin{equation}
\label{eq:tgsm_accumulated_angle}
\Theta_{\rm TGSM}(z)
=
\mu_0\Lambda_0
\arctan \left(
\frac{z}{\Lambda_0}
\right),
\end{equation}
is a scalar measure of the hidden rotational transport.
Note that the accumulated angular displacement is independent of the launch radius and saturates because the angular rate $\mu(z)$ decreases during propagation.
Although the intensity remains Gaussian, different values or signs of \(\mu_0\) produce different spiral flow topologies. The transverse curves are expanding spirals,
\begin{equation}
\label{eq:tgsm_cartesian_trajectories}
x(z)=r(z)\cos\varphi(z),
\qquad
y(z)=r(z)\sin\varphi(z).
\end{equation}
For \(z\to\infty\), the total accumulated twist angle is
\begin{equation}
\label{eq:tgsm_asymptotic_angle}
\Delta\varphi_\infty
=
\frac{\pi}{2}\mu_0\Lambda_0 .
\end{equation}
The sign of \(\mu_0\) fixes the handedness of the trajectories.
The signs of the curvature and twist contributions follow the time-harmonic and propagation conventions specified in Sec.~\ref{sec:2.1}.

This example separates intensity from transport in the simplest possible way: the intensity is Gaussian and rotationally symmetric, whereas the generalized flux contains a twist-controlled circulation and vorticity. The hidden topology is encoded entirely in the two-point phase of the CSD.

\subsection{Laguerre-Christoffel-Darboux beams}

LCD beams provide a second, and in some sense sharper, demonstration of hidden flow topology. In the TGSM case, the twist produces rotational transport while the intensity remains Gaussian. In the LCD case, two distinct partially coherent sources can be constructed with the same radial intensity profile but different angular coherence structure. Their intensities are identical, yet their generalized fluxes and trajectories are not.

LCD beams are built as finite incoherent superpositions of Laguerre-Gaussian modes~\cite{Siegman1986} with fixed topological charge and suitable modal weights. As a consequence of the Christoffel-Darboux identity, their CSD can be expressed in closed form for any number of radial modes. The complete CSD is shape-invariant under paraxial propagation apart from transverse scaling and phase curvature~\cite{Martinez-Herrero:21,MartinezHerrero2021Photonics}. Here we use this family to show explicitly that equality of intensity does not imply equality of transverse transport.

We take as starting point the propagated single-charge LCD CSD, written directly in terms of the transverse polar coordinates \(\mathbf r_j=(r_j,\varphi_j)\):
\begin{multline}
\label{eq:lcd_csd}
\mathcal{W}_{N,m}(\mathbf r_1,\mathbf r_2,z)
=
A(z)
\exp\left[
-\frac{r_1^2+r_2^2}{w^2(z)}
\right]
\exp\left[
-\frac{ik(r_1^2-r_2^2)}{2R(z)}
\right]
\\
\times
\exp[-im(\varphi_1-\varphi_2)]
\left(
\frac{2r_1r_2}{w^2(z)}
\right)^m
G_{N,m}(r_1,r_2;z) ,
\end{multline}
where $m$ here denotes the modulus of the topological charge (later, the explicit presence of a sign will indicate whether the topological charge is positive or negative).
Here $A(z)$ is fixed by the total spectral power $P$. Using the Laguerre orthogonality relation and imposing \[ P=\int I_{N,m}(r,z)\,d^2{\bf r}, \] one obtains \[ A(z)=\frac{2P}{\pi w^2(z)(N+1)}. \] This normalization ensures conservation of the total spectral power during paraxial propagation.
The width and radius of curvature are the usual Laguerre-Gaussian propagation parameters,
\begin{equation}
\label{eq:lcd_propagation_parameters}
w(z) = w_0 \ \! \sqrt{ 1+\frac{z^2}{z_R^2} },
\qquad
R(z)
=
z
\left(
1+\frac{z_R^2}{z^2}
\right),
\end{equation}
with Rayleigh range
\begin{equation}
\label{eq:lcd_rayleigh_range}
z_R
=
\frac{kw_0^2}{2}.
\end{equation}
The first exponential in Eq.~\eqref{eq:lcd_csd} gives the Gaussian envelope of the two-point correlation. The second exponential is the spherical curvature phase. The factor \(\exp[-im(\varphi_1-\varphi_2)]\) contains the vortex-like angular correlation associated with the fixed topological charge \(m\). The algebraic factor is inherited from the azimuthal order of the underlying Laguerre-Gaussian modes. The remaining radial coherence structure is contained in \(G_{N,m}\).

For this LCD source,
\begin{equation}
\label{eq:lcd_radial_kernel}
G_{N,m}(r_1,r_2;z)
=
\sum_{n=0}^{N}
\frac{n!}{(n+m)!}
L_n^m
\left(
\frac{2r_1^2}{w^2(z)}
\right)
L_n^m
\left(
\frac{2r_2^2}{w^2(z)}
\right).
\end{equation}
Thus \(N+1\) radial Laguerre-Gaussian modes contribute incoherently, all with the same azimuthal index \(m\). The Gouy phases of the individual modes cancel within each modal contribution to the CSD, which is why they do not appear explicitly in Eq.~\eqref{eq:lcd_csd}. The propagation is encoded through \(w(z)\) and \(R(z)\).

The key mathematical property of LCD beams is that the finite sum in Eq.~\eqref{eq:lcd_radial_kernel} can be evaluated in closed form by the Christoffel-Darboux identity.
Let us define the dimensionless variables
\begin{equation}
\label{eq:lcd_chi}
\chi_j \equiv \frac{2r_j^2}{w^2(z)} , \qquad j = 1, 2.
\end{equation}
For \(r_1\neq r_2\), one has
\begin{equation}
\label{eq:lcd_christoffel_darboux}
G_{N,m}(r_1,r_2;z) =
\frac{(N+1)!}{(N+m)!}\frac{Q_{N,m}(\chi_1,\chi_2)}{\chi_1 - \chi_2},
\end{equation}
with
\begin{equation}
\label{eq:lcd_Q}
Q_{N,m} \equiv L_N^m(\chi_1)L_{N+1}^m(\chi_2) - L_{N+1}^m(\chi_1)L_N^m(\chi_2).
\end{equation}
This is the reason why LCD beams remain analytically convenient even when many radial modes are included. At coincidence points, the intensity is obtained from the diagonal of the CSD, so we have
\begin{equation}
\label{eq:lcd_intensity}
I_{N,m}(r,z)
=
A(z)
\exp\left[
-\frac{2r^2}{w^2(z)}
\right]
\left(
\frac{2r^2}{w^2(z)}
\right)^m
G_{N,m}(r,r;z) ,
\end{equation}
where
\begin{equation}
\label{eq:lcd_christoffel_darboux2}
G_{N,m}(r,r;z) = \frac{(N+1)!}{(N+m)!}
 \left[
 L_{N+1}^m(\chi)\, \frac{dL_N^m(\chi)}{d\chi} - L_N^m(\chi)\, \frac{dL_{N+1}^m(\chi)}{d\chi}
 \right] , \quad \chi \equiv \frac{2r^2}{w^2(z)} .
\end{equation}
For \(m\neq 0\), the intensity has a donut-like profile. The radius of the central dark region increases with \(m\) and decreases with \(N\), while the overall profile remains shape-invariant under paraxial propagation apart from the scale factor \(w(z)/w_0\)~\cite{MartinezHerrero2021Photonics}. The important point for the present discussion is that this radial intensity profile does not determine the angular part of the generalized flux. The latter depends on the phase structure of the CSD, and hence on the angular coherence content of the source.

The transverse spectral flux follows immediately from the modal structure. All coherent modes entering the single-charge LCD sum have the same transverse phase gradient,
\begin{equation}
\label{eq:lcd_phase_gradient}
\nabla_\perp
\left(
\frac{kr^2}{2R(z)}
+
m\varphi
\right)
=
\frac{kr}{R(z)}\hat{\mathbf r}
+
\frac{m}{r}\hat{\boldsymbol\varphi}.
\end{equation}
The radial amplitudes and Laguerre polynomials contribute only real terms to \(\Phi^*\nabla_\perp\Phi\), and therefore do not enter the imaginary part.
Consequently, with the phase convention adopted in Eq.~\eqref{eq:lcd_csd}, the following single-charge results correspond to the positive charge $+m$, starting from
\begin{equation}
\label{eq:lcd_flux_single_charge}
{\mathbf J}_{\perp,N}^{(m)}(\mathbf r,z)
=
I_{N,m}(r,z)
\left[
\frac{r}{R(z)}\hat{\mathbf r}
+
\frac{m}{kr}\hat{\boldsymbol\varphi}
\right].
\end{equation}
The normalized flow is therefore universal within the single-charge family:
\begin{equation}
\label{eq:lcd_normalized_flow_single_charge}
\frac{{\mathbf J}_{\perp,N}^{(m)}(\mathbf r,z)}{I_{N,m}(r,z)} =
\frac{r}{R(z)}\hat{\mathbf r} + \frac{m}{kr}\hat{\boldsymbol\varphi}.
\end{equation}
This result provides a complementary example of the discussion in Sec.~\ref{sec:IID}. Increasing the number of mutually incoherent radial modes changes the spectral-density profile through $G_{N,m}$, but it does not modify the normalized flow because all modes carry the same topological charge. From Eq.~\eqref{eq:lcd_normalized_flow_single_charge}, its radial and azimuthal components are
\begin{equation}
\label{eq:lcd_velocity_components_single_charge}
v_r(r,z)=\frac{r}{R(z)},
\qquad
v_\varphi(r,z)=\frac{m}{kr}.
\end{equation}
For $m\neq0$, the spectral density vanishes on the propagation axis and the azimuthal velocity is singular at $r=0$. The regular velocity field and its streamlines are therefore defined on the punctured transverse plane $r>0$. The corresponding flux remains well behaved at the core for the LCD profiles considered here because the spectral density vanishes sufficiently rapidly as $r\to0$.

The circulation around a circle of radius \(r\) is
\begin{equation}
\label{eq:lcd_circulation_single_charge}
\Gamma_m(r,z)
=
2\pi r\,v_\varphi(r,z)
=
\frac{2\pi m}{k},
\end{equation}
which is independent of \(r\) and \(z\). Away from the beam axis, the corresponding vorticity vanishes,
\begin{equation}
\label{eq:lcd_vorticity_single_charge}
\Omega_m(r,z)
=
\frac{1}{r}
\frac{\partial}{\partial r}
\left(
r\frac{m}{kr}
\right)
=
0,
\qquad r\neq 0,
\end{equation}
while the nonzero circulation reflects the singular angular phase structure associated with the topological charge.

This behavior differs from the distributed vorticity of the TGSM beam in Sec.~4.1. Whereas $v_\varphi=\mu(z)r$ gives the uniform vorticity $\Omega_{\rm TGSM}=2\mu(z)$, the single-charge LCD flow has zero classical vorticity on the punctured plane $r>0$ but nonzero circulation $\Gamma_m=2\pi m/k$ around the singular core. In a distributional description, its vorticity is concentrated at that core.

The radial modal complexity of the LCD source is therefore contained in the scalar intensity, whereas the hidden angular transport is controlled by \(m\).
Note that, for a fixed absolute topological charge $m$, reversing the sign of the charge leaves the spectral density unchanged but reverses the azimuthal component of the generalized velocity:
\begin{align}
 I_N^{(+m)}(r,z) & = I_N^{(-m)}(r,z) = I_{N,m}(r,z), \\
 v_{\varphi,N}^{(+m)}(r,z) &= -v_{\varphi,N}^{(-m)}(r,z) = v_\varphi(r,z) .
\end{align}
Accordingly,
\begin{equation}
 \Gamma_{+m}=-\Gamma_{-m} = \Gamma_m,
\end{equation}
and the fixed-plane streamlines have the opposite handedness.
Thus, for a fixed $m$, the single-charge $-m$ source contains no additional spectral-density information with respect to the $+m$ source; it differs only in the reversed sense of the azimuthal flow.

The associated trajectories satisfy
\begin{equation}
\label{eq:lcd_trajectory_equations_single_charge}
\frac{dr}{dz}
=
\frac{r}{R(z)},
\qquad
\frac{d\varphi}{dz}
=
\frac{m}{kr^2}.
\end{equation}
Using Eq.~\eqref{eq:lcd_propagation_parameters}, the radial equation integrates to
\begin{equation}
\label{eq:lcd_radial_solution}
r(z)
=
r_0
\frac{w(z)}{w_0}
=
r_0
\left(
1+\frac{z^2}{z_R^2}
\right)^{1/2}.
\end{equation}
Substitution into the angular equation gives
\begin{equation}
\label{eq:lcd_angular_solution}
\varphi(z)
=
\varphi_0
+
\frac{mw_0^2}{2r_0^2}
\arctan \left(
\frac{z}{z_R}
\right).
\end{equation}
The accumulated angular displacement is
\begin{equation}
\label{eq:lcd_accumulated_angle}
\Theta_m(z;r_0)
=
\frac{mw_0^2}{2r_0^2}
\arctan \left(
\frac{z}{z_R}
\right).
\end{equation}
Unlike the TGSM case, this rotation depends on the initial radius \(r_0\). Inner trajectories rotate more strongly than outer ones, reflecting the \(1/r\)-dependence of the azimuthal velocity associated with the angular phase structure. Thus a single-charge LCD source generates spiral average trajectories. The radial part is the usual diffractive scaling of a Laguerre-Gaussian beam, whereas the azimuthal part is controlled by the topological charge \(m\).

The same LCD construction can be modified by assigning equal weights to modes with charges \(+m\) and \(-m\), which leads to a \textit{balanced} CSD,
\begin{equation}
 \mathcal{W}_N^{(\rm bal)} = \frac{1}{2} \left[ \mathcal{W}_N^{(+m)} + \mathcal{W}_N^{(-m)} \right] .
\end{equation}
In this case, the intensity remains the same as in the single-charge case, i.e., $I_N^{(\rm bal)} = I_{N,m}$, while the angular dependence of the degree of coherence changes from a vortex-like phase factor to a cosine modulation:
\begin{equation}
\label{eq:lcd_degree_coherence_balanced}
\gamma_N^{(\rm bal)} (\mathbf r_1,\mathbf r_2, z) = \frac{G_{N,m}(r_1,r_2;z) \cos[m(\varphi_2-\varphi_1)]}{\sqrt{G_{N,m}(r_1,r_1;z) G_{N,m}(r_2,r_2;z)}} \exp[-ik(r_1^2-r_2^2)/2R(z)] .
\end{equation}
The two modal families carry opposite azimuthal fluxes, which cancel exactly. The radial contribution is the same for both families, and therefore
\begin{equation}
\label{eq:lcd_flux_balanced}
\mathbf J_{\perp,N}^{(\rm bal)}(\mathbf r,z) = I_{N,m}(r,z) \frac{r}{R(z)} \hat{\mathbf r}.
\end{equation}
Thus, as anticipated in Sec.~\ref{sec:IID}, coherence affects the transport topology through the specific modal and phase structure of the CSD: unlike the incoherent addition of radial modes with a common charge, the balanced combination of opposite azimuthal momenta cancels the azimuthal component of the normalized flow.
The trajectories are purely radial:
\begin{equation}
\label{eq:lcd_trajectory_equations_balanced}
\frac{dr}{dz}
=
\frac{r}{R(z)},
\qquad
\frac{d\varphi}{dz}
=
0.
\end{equation}
The two LCD sources therefore satisfy
\begin{equation}
\label{eq:lcd_equal_intensity}
I_N^{(+m)}(r,z) = I_N^{(\rm bal)} (r,z) = I_{N,m}(r,z) ,
\end{equation}
but
\begin{equation}
\label{eq:lcd_unequal_flux}
\mathbf J_{\perp,N}^{(+m)}(\mathbf r,z) \neq
  \mathbf J_{\perp,N}^{(\mathrm{bal})}(\mathbf r,z).
\end{equation}
In particular,
\begin{equation}
\label{eq:lcd_circulation_comparison}
\Gamma_{+m}(r,z) = \frac{2\pi m}{k}, \qquad \Gamma_{\rm bal}(r,z) = 0 .
\end{equation}
For these two equal-intensity LCD sources, the flow contrast defined in Eq.~\eqref{eq:flow_contrast} can be evaluated directly. The single-charge and balanced sources have, respectively,
\[
 {\bf v}_{\perp,N}^{(+m)}(\mathbf r,z) = \frac{r}{R(z)}\hat{\bf r} + \frac{m}{kr}\hat{\boldsymbol\varphi}, \qquad {\bf v}_{\perp,N}^{(\rm bal)}(\mathbf r,z) = \frac{r}{R(z)}\hat{\bf r},
\]
and hence
\[
 {\cal C}_{+m,\rm bal}^{\,2}(z) = \frac{ \displaystyle \frac{m^2}{k^2} \int I_{N,m}(r,z)\,r^{-2}\,d^2{\bf r} }{ \displaystyle \frac{2}{R^2(z)} \int I_{N,m}(r,z)\,r^2\,d^2{\bf r} + \frac{m^2}{k^2} \int I_{N,m}(r,z)\,r^{-2}\,d^2{\bf r} }.
\]
The contrast is finite for $m\neq0$, because the central zero of the LCD spectral density regularizes the $r^{-2}$ contribution, while the Gaussian envelope ensures convergence at large radius. At the waist, $R(0)\rightarrow\infty$, the common radial velocity vanishes and ${\cal C}_{+m,\rm bal}(0)=1$. During propagation, the common radial expansion contributes to the denominator and reduces the relative contrast associated with the azimuthal flow.

\begin{figure*}[!t]
\centering
\includegraphics[width=\textwidth]{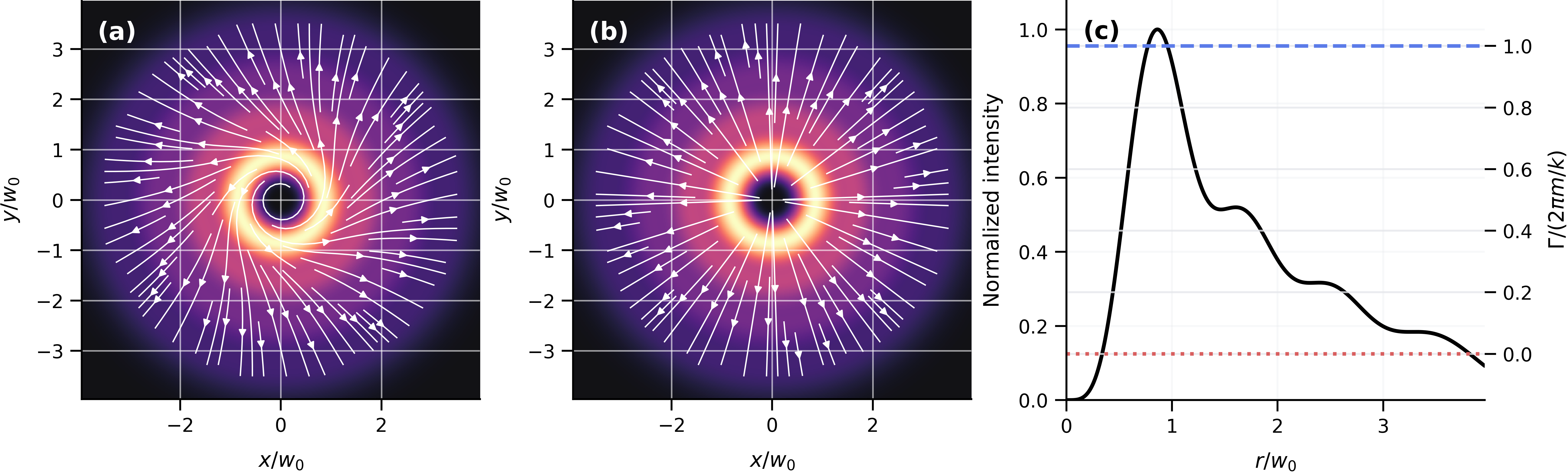}
\caption{
Hidden transverse-flow topology in LCD beams. The background color represents the normalized spectral density $I_{N,m}(r,z)$ at the fixed propagation plane $z=z_R$, which is identical in panels (a) and (b).
White curves are fixed-plane streamlines of the generalized transverse velocity field at $z=z_R$.
Their displayed extent is determined by the common seed distribution, plotting domain, and numerical termination criteria and therefore does not represent a propagation distance or transport time.
(a) A single-charge LCD source with topological charge $+m$ displays spiral fixed-plane streamlines because $v_\varphi=m/(kr)$.
The corresponding single-charge $-m$ case has the same spectral-density distribution and the opposite azimuthal velocity and circulation; its streamlines therefore have the opposite handedness.
(b) A balanced incoherent superposition of charges $+m$ and $-m$ has the same spectral-density profile, but the azimuthal flux cancels and the flow is purely radial.
The same seed points, placed within the bright annular region of the spectral-density distribution, and identical plotting and termination settings are used in panels (a) and (b).
(c) Radial spectral-density profile and circulation normalized to $2\pi m/k$. The single-charge source has $\Gamma_{+m}=2\pi m/k$, whereas the balanced source has $\Gamma_{\rm bal}=0$.
Parameters: $N=3$, $m=2$, $k w_0=20$, and $z/z_R=1$, with $z_R=kw_0^2/2$.
}
\label{fig:lcd_hidden_flow}
\end{figure*}

The contrast between the two cases is summarized visually in Fig.~\ref{fig:lcd_hidden_flow} through fixed-plane streamlines of the corresponding velocity fields at $z=z_R$.
The same spectral-density distribution supports two inequivalent velocity fields: a spiral flow with nonzero circulation for the single-charge source and a purely radial flow with zero circulation for the balanced opposite-charge source.
The streamline seeds are used only to visualize the local flow topology at the selected plane and should not be interpreted as initial conditions for propagation trajectories.

Thus the two beams are indistinguishable by their intensity profiles but dynamically inequivalent. In the single-charge case, the CSD carries a vortex-like angular phase and the generalized trajectories spiral. In the balanced opposite-charge case, the same intensity is accompanied by a real cosine angular coherence factor, the azimuthal flux cancels, and the trajectories remain radial.

This example is the most direct realization of the hidden-flow-topology principle. The intensity fixes where the averaged optical energy is located, but not how it moves. The latter information is contained in the off-diagonal phase structure of the CSD and becomes visible only through the generalized flux.

The two examples reveal two different mechanisms by which hidden flow topology can arise. In TGSM beams, a nonseparable twist phase produces a distributed azimuthal velocity \(v_\varphi=\mu(z)r\), giving a uniform vorticity \(\Omega_{\rm TGSM}=2\mu(z)\) under a Gaussian intensity envelope. In single-charge LCD beams, the angular coherence structure produces \(v_\varphi=m/(kr)\), giving a radius-independent circulation \(\Gamma_m=2\pi m/k\). In balanced opposite-charge LCD beams, the same intensity profile is retained but the opposite azimuthal fluxes cancel. These results show that intensity, coherence magnitude, and flow topology are distinct layers of information contained in the CSD.


\section{Discussion}
\label{sec:discussion}

The results above show that the intensity distribution is not, by itself, a complete descriptor of transverse optical transport in partially coherent structured light. The spectral density specifies where the averaged optical energy is located, but it does not determine how this energy is locally redistributed during propagation. That information is contained in the generalized flux extracted from the CSD. The hidden transverse-flow topology is therefore the topology of the velocity field \(\mathbf v_\perp=\mathbf J/I\) beyond what can be inferred from \(I(\mathbf r,z)\) alone.

This distinction is especially relevant when the CSD contains nontrivial phase correlations. In a fully coherent beam, transverse transport is governed by the gradient of a single optical phase. In a partially coherent beam, such a phase is no longer available. Nevertheless, the two-point correlation phase can still carry transverse-momentum information.
Equation~\eqref{eq:generalized_flux_csd} extracts the antisymmetric phase-gradient content of the CSD at coincidence points and converts it into a local flux.
In this sense, the generalized flux is not a secondary quantity derived from intensity; it is an independent layer of physical information encoded in the CSD.

The TGSM example illustrates this point in its simplest rotational form. The spectral density remains a circular Gaussian, and therefore contains no visible signature of angular transport. However, the nonseparable twist phase produces an azimuthal velocity \(v_\varphi=\mu(z)r\), a circulation \(\Gamma_{\rm TGSM}=2\pi\mu(z)r^2\), and a uniform vorticity \(\Omega_{\rm TGSM}=2\mu(z)\). Thus, the twist is revealed not as a deformation of the intensity profile, but as a distributed rotational structure of the generalized flow.

\begin{figure*}[!t]
\centering
\includegraphics[width=\textwidth]{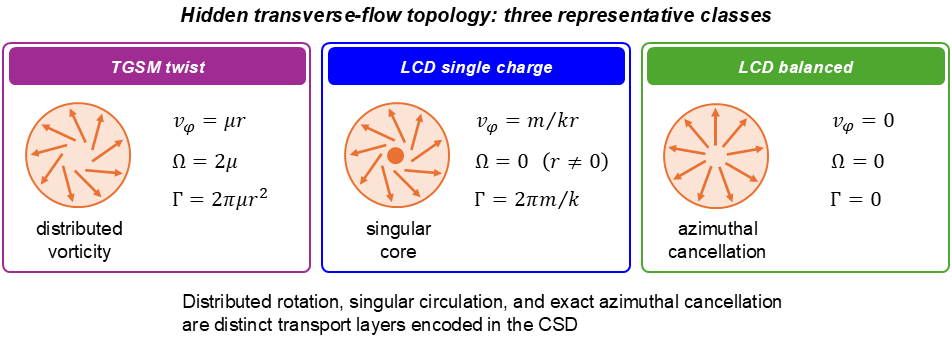}
\caption{
Schematic classification of hidden transverse-flow topology in the analytical cases considered in this work.
A TGSM beam with twist displays distributed rotational transport, with \(v_\varphi=\mu r\), circulation \(\Gamma=2\pi\mu r^2\), and uniform vorticity \(\Omega=2\mu\).
A single-charge LCD beam displays angular transport of topological origin, with \(v_\varphi=m/(kr)\) and radius-independent circulation \(\Gamma=2\pi m/k\), while the vorticity vanishes away from the singular core.
A balanced LCD source with opposite charges \(+m\) and \(-m\) has the same radial intensity profile as the single-charge case, but the azimuthal flux cancels exactly, giving \(v_\varphi=0\), \(\Gamma=0\), and \(\Omega=0\).
}
\label{fig:topology_diagnostics_summary}
\end{figure*}

The LCD example provides an even sharper demonstration. A single-charge LCD source and a balanced opposite-charge LCD source may share the same radial intensity profile. They are therefore intensity-equivalent but transport-inequivalent. At the flux level, the single-charge source carries a nonzero circulation \(2\pi m/k\) and produces spiral trajectories, whereas the balanced source cancels the opposite azimuthal fluxes and produces purely radial trajectories.

The azimuthal generalized flux is also related to the ensemble-averaged $z$-component of the OAM density associated with the scalar paraxial current.
Within the present normalization, this current-based OAM density is proportional to $({\bf r}\times{\bf J})_z=rJ_\varphi$.
For the TGSM beam, $rJ_\varphi=\mu(z)r^2S(r,z)$, with a sign fixed by the twist parameter~\cite{Wang2022Nanophotonics}. For the single-charge LCD source, $rJ_\varphi=m I_{N,m}(r,z)/k$, whereas in the balanced opposite-charge source the ensemble-averaged azimuthal flux, and hence the corresponding OAM density, cancels. This current-based quantity characterizes the ensemble-averaged OAM associated with the generalized transverse flux, but it does not determine the complete OAM spectrum or its fluctuations.

The formulation also clarifies the role of partial coherence in structured light. Partial coherence is often described as a mechanism that reduces visibility, smooths fine spatial features, or redistributes modal weights. These effects are important, but they do not exhaust the physical consequences of coherence engineering.
By modifying the off-diagonal phase structure of the CSD, partial coherence can also create, suppress, or transform transverse-flow topology.
In particular, coherence phases may generate transport structures that have no direct counterpart in the spectral density.
Related recent work has also shown that nontrivial topological structures can be encoded in coherence singularities of fluctuating optical fields, rather than in the average intensity alone~\cite{Wang2025CoherenceKnots}.

The trajectory representation is therefore not merely a visualization device. It provides a way to classify partially coherent beams according to their effective transport dynamics. Circulation, vorticity, accumulated angular displacement, and flow inequivalence for equal intensities are examples of diagnostics that distinguish beams with similar or identical intensity profiles. Such quantities may be especially useful for partially coherent vortex beams, twisted Schell-model sources, and structured-coherence fields designed through modal or correlation engineering.
The three cases analyzed above can therefore be viewed as representative classes of hidden transverse-flow topology, as summarized in Fig.~\ref{fig:topology_diagnostics_summary}. 
They differ not by their intensity information alone, but by how the off-diagonal phase structure of the CSD organizes azimuthal transport.

This classification suggests that the proposed diagnostics may be useful in situations where different sources or propagation channels produce similar intensity patterns but different correlation phases.
Examples include coherence-engineered vortex beams, partially coherent beams carrying OAM, propagation through random media, and source design for robust structured-light transport.
In this context, recent experiments have shown that topological-charge information can be encoded in the spatial-coherence structure of partially coherent vortex beams and recovered even when intensity-based signatures are degraded~\cite{Zhang2024RobustOAM}.
In such settings, intensity-based criteria may fail to distinguish beams with different local momentum content, whereas circulation, vorticity, and flow inequivalence provide transport-sensitive figures of merit. The approach may also help in designing partially coherent fields whose intensity is constrained by an application while their transverse transport is tailored through the off-diagonal structure of the CSD.

Fields with hard apertures or non-smooth source correlations may fail to satisfy the pointwise differentiability assumption at the source plane, particularly at aperture edges. The present formulation then applies only away from those singular sets or after propagation to a plane where the CSD has acquired the required regularity. A formulation based on weak or distributional derivatives may be possible, but it lies beyond the scope of the present work.

The generalized-flux construction is not tied to the diffractively expanding, self-similar examples considered here. It applies locally to any physically admissible paraxial CSD that is sufficiently differentiable at coincidence. Ideal nondiffracting partially coherent fields are included at this local level, although they should be understood as idealized infinite-energy models. Finite-energy realizations require truncation or apodization and are only quasi-nondiffracting over a finite propagation range. Non-smooth truncation edges are subject to the differentiability limitations discussed above.

Propagation invariance of the spectral density does not require the transverse flux to vanish. According to the continuity equation, Eq.~\eqref{eq:continuity_equation}, a $z$-invariant spectral density only requires $\nabla_\perp\cdot{\bf J}_\perp=0$, rather than ${\bf J}_\perp=0$. Therefore, a propagation-invariant CSD may sustain a stationary azimuthal current that remains hidden in its invariant spectral-density profile. For a regular axisymmetric field with $J_r=0$ and $J_\varphi\neq0$, the fixed-plane transverse streamlines are closed circles, whereas the corresponding three-dimensional propagation trajectories wind helically around the propagation axis while remaining at constant radius.
Incoherent mixtures of nondiffracting modes carrying opposite azimuthal currents may additionally produce counter-rotating radial domains or cancel the mean azimuthal flux. Since the spectral density contains the sum of the modal intensities whereas the azimuthal flux retains the signs of the modal currents, different CSDs may support the same propagation-invariant spectral density while exhibiting different transport topologies.
A detailed analysis of these configurations lies beyond the scope of the present work.

Although the present work is theoretical, the generalized flux is, in principle, reconstructible from sufficiently resolved phase-sensitive measurements of the complex CSD. Introducing the midpoint and separation coordinates ${\bf r}=({\bf r}_1+{\bf r}_2)/2$ and ${\boldsymbol\Delta}={\bf r}_2-{\bf r}_1$, Eq.~\eqref{eq:generalized_flux_csd} becomes
\begin{equation}
{\bf J}({\bf r},z) = \frac{1}{ik} \left. \nabla_{\boldsymbol\Delta} W\left({\bf r}-\frac{{\boldsymbol\Delta}}{2}, {\bf r}+\frac{{\boldsymbol\Delta}}{2},z\right) \right|_{{\boldsymbol\Delta}=0}.
\end{equation}
Thus, reconstruction of the local flux requires complex CSD data in a neighborhood of the coincidence set, rather than over the complete four-dimensional transverse domain. In practice, the derivative could be estimated from measurements at small opposite separations by centered finite differences or a local polynomial fit. The measured spectral density then gives ${\bf v}_\perp={\bf J}/I$ only in regions where the signal is sufficiently above the noise floor.

Several experimental strategies may provide the required complex second-order coherence data. Generalized Hanbury Brown--Twiss schemes with coherent reference fields can retrieve the real and imaginary parts of the complex degree of spatial coherence, while phase-space tomography and incoherent modal decomposition provide alternative reconstructions of the mutual intensity or CSD~\cite{Huang2020,Tian2012,Lu2023}. These approaches have different statistical assumptions, sampling requirements, and experimental limitations and are therefore not interchangeable in all regimes.

At a more general level, self-configuring photonic networks provide a promising route for analyzing, processing, and generating partially coherent light through a modal decomposition of its coherence matrix ~\cite{RoquesCarmes2024PCLA}. Such architectures can separate the field into mutually incoherent orthogonal components and thereby recover a discrete modal representation of its coherence matrix. Applied to the present problem, this representation would still have to provide sufficient spatial sampling near the CSD diagonal for the coincidence derivative, and hence the generalized flux, to be reconstructed.

In all these approaches, reconstruction of the generalized flux would be sensitive to phase stability, transverse sampling, detector resolution, calibration errors, statistical convergence, and the amplification of measurement noise by numerical differentiation. A practical implementation would therefore require local fitting or regularization, uncertainty propagation, and an intensity or signal-to-noise threshold before evaluating the velocity field and integrating its trajectories. The required $z$ dependence could be obtained either from CSD measurements at several propagation planes or, when the propagation operator is accurately known, by numerical propagation from a reconstructed input-plane CSD. The trajectories should therefore be understood as reconstructed integral curves of an estimated correlation-induced transport field, rather than as directly measured material paths of individual photons or random field realizations.

Finally, the formal analogy with reduced quantum dynamics is worth emphasizing~\cite{Sanz_EPJD:07,Sanz_CPL:09}. The paraxial and Schr\"odinger equations share the same mathematical structure, and the CSD plays a role analogous to a density matrix in the transverse-position representation. This analogy clarifies why the generalized flux must be constructed from the complete statistical operator: the total current is the weighted sum of the modal currents, whereas the effective velocity is the ratio of the total current to the total spectral density and is therefore not, in general, a direct average of the individual modal velocities.

The analogy is nevertheless formal. In the present optical problem, $z$ is the paraxial propagation coordinate, the CSD describes classical second-order field correlations, and the normalization and experimental interpretation concern optical energy transport rather than quantum probability. At a conceptual level, this connection is related to previous reconstructions of average optical trajectories from local momentum measurements~\cite{Kocsis2011,Bliokh2013}. The present construction differs, however, in that the trajectories are defined at the level of second-order field correlations, rather than from a pure coherent field or a postselected single-photon ensemble.


\section{Conclusions}
\label{sec:conclusions}

We have developed a trajectory-based formulation for partially coherent structured light that reveals transverse-flow information hidden in the CSD. Starting from the continuity equation for the averaged intensity, we defined a generalized transverse flux and the corresponding effective velocity field.
The resulting trajectories reduce to the usual coherent flux trajectories in the single-mode limit and remain well defined for physically admissible quasimonochromatic partially coherent fields whose CSD is sufficiently differentiable at coincidence and whose spectral density is nonzero in the region considered.

The central result is that the CSD encodes not only intensity and coherence, but also an effective transport topology. This topology can remain invisible at the level of the spectral density. By extracting the antisymmetric phase-gradient content of the CSD at coincidence points, the generalized flux converts two-point phase correlations into local transverse energy flow.

Two analytical beam families demonstrate this principle.
In TGSM beams, a nonseparable coherence phase produces a distributed azimuthal flow, nonzero circulation, and uniform vorticity, while the intensity remains Gaussian. In LCD beams, two sources can have the same radial intensity profile but different generalized fluxes: a single-charge source produces spiral trajectories, whereas a balanced opposite-charge source produces purely radial ones.

These results support the main conclusion of this work: partial coherence should not be viewed only as a loss of contrast or a smoothing mechanism. It can also reorganize the hidden topology of transverse energy flow. Generalized flux trajectories therefore provide a direct way to uncover, classify, and quantify correlation-induced transport structures that remain hidden in intensity-based descriptions.





\section*{Disclosures}
The authors declare no conflicts of interest.




\end{document}